\documentclass[10pt,final,journal,twoside]{IEEEtran}
\usepackage{url}
\usepackage{float}
\usepackage[mathscr]{eucal}
\usepackage{ifpdf}
\usepackage{cite}
\usepackage{graphicx}
\usepackage[cmex10]{amsmath}
\usepackage{amssymb}
\usepackage{algorithmic}
\usepackage{units}
\usepackage{setspace}
\usepackage{algorithm}
\usepackage{array}
\usepackage{amsthm}
\usepackage{diagbox}
\usepackage{multirow}
\usepackage{enumerate}
\usepackage{color}
\usepackage{mathtools}
\newcommand{\subparagraph}{}
\usepackage[compact]{titlesec}
\usepackage{subcaption}

\usepackage{lmodern}
\usepackage{graphics}
\usepackage{chngcntr}
\usepackage{tikz}
\usetikzlibrary{arrows,calc}

\usepackage{subcaption}
\usepackage{wrapfig,booktabs}
\begin{document}
\title{Computational Models of Human Decision-Making With Application to the Internet of Everything}
\author{
\IEEEauthorblockN{Setareh Maghsudi and Max Davy}
\thanks{S. Maghsudi is with the Department of Electrical Engineering and Computer Science, Technical University of Berlin, 10623 Berlin, Germany (email: maghsudi@tu-berlin.de). M. Davy is with the School of Mathematics and Statistics, University of Sydney, Camperdown NSW 2006, Australia (email: mdav9030@uni.sydney.edu.au).}
}
\maketitle
\begin{abstract}
The concept of the \textit{Internet of Things} (IoT) first appeared a few decades ago. Today, by the ubiquitous wireless connectivity, the boost of machine learning and artificial intelligence, and the advances in big data analytics, it is safe to say that IoT has evolved to a new concept called the \textit{Internet of Everything} (IoE) or the \textit{Internet of All}. IoE has four pillars: Things, human, data, and processes, which render it as an inhomogeneous large-scale network. A crucial challenge of such a network is to develop management, analysis, and optimization policies that besides utility-maximizer machines, also take irrational humans into account. We discuss several networking applications in which appropriate modeling of human decision-making is vital. We then provide a brief review of computational models of human decision-making. Based on one such model, we develop a solution for a task offloading problem in fog computing and we analyze the implications of including humans in the loop. 
\end{abstract}
\section{Introduction}
\label{sec:Intro}
Integrating the human element in digital technology is a crucial aspect of digitalization that blurs the boundaries between the cyber and physical worlds. Many emerging application domains of IoE, such as intelligent transportation systems and autonomous driving include humans in addition to machines. An example of the potential of the human-in-the-loop can be seen on the basis of the following assumption: Given the great dissemination of smart-phones as a result of their vast functionality and wide price-range, near-future human-in-the-loop systems will be heavily based on smart-phone technology. Consequently, their powerful computation and sensing capabilities can be utilized to address networking challenges such as scarcity of resources. 

While combining humans and machines to achieve sustainable resource allocation and efficient service provision in networks is promising, the integration of humans and machines in a unique network is challenging due to the following reasons: (i) Humans make mistakes, often due to inaccurate beliefs and imprecise predictions; (ii) Humans often act irrationally and based on heuristics; (iii) Humans think and act in different manners as a result of their unique background, including personality and experiences. These characteristics stand in contrast to the most common assumptions of rational decision-making and strategic behavior. As a result, human behavior cannot be formalized by using conventional models such as simple concave utility functions, and new models need to be developed. For example, the well-being of humans is strongly based on their utility compared to that of others, rather than the absolute magnitude of the utility on its own. Another example is the herding phenomenon observed in humans, which describes how individuals in a group can act collectively without centralized direction. A further example is the information-avoidance behavior, where, by consciously ignoring the available information, individuals perform some tasks that oppose their values \cite{Samson18:TBG}. For the analysis, management, and optimization of networks with both humans and machines, appropriate modeling of this irrational human behavior is crucial.

Moreover, the presence of the human often harms the reliability of the data based on which the network is controlled. Often suppressed by norm-related conflicts, information externalities, and emotions, self-reported human data is severely biased. Therefore, learning or predicting the human's intent is highly complicated. 

In Section \ref{sec:ASoA}, we show that considering human-introduced characteristics for network management by using appropriate models is needed, and we review some research papers that take the human into account when addressing the optimization problem. In Section \ref{sec:HumanModel}, we provide a brief survey on models for human behavior. In Section \ref{sec:DevModel}, we describe some important issues in developing models. In Section \ref{sec:Example}, we provide an exemplary application of such modeling in edge computing. Section \ref{sec:Conc} concludes the paper.
\section{State-of-the-Art and Applications}
\label{sec:ASoA}
In this section, we first briefly review some relevant literature. Afterward, we describe some applications of agent-based modeling of human behavior. 
\subsection{State-of-the-Art}
\label{sec:SoA}
Human-in-the-loop systems can be discussed from several perspectives. Article \cite{Schirner13:FHL} is a forward-looking survey on human-in-the-loop control in cyber-physical systems, with an emphasis on robotics and human-machine interaction. A theoretical framework for shared control between humans and autonomous agents is proposed in \cite{Gopinath17:HiL}, which is based on nonlinear optimization. Architectural issues are also a concern in designing socially-intelligent systems. For example, \cite{Wood08:SenQ} is an attempt for including wireless sensor networks into the human-in-the-loop cyber-physical system. In this work, the network is designed to support user-driven applications, through peer-to-peer in-network queries between resource-constrained devices. Another important topic, which is the focus of this paper, is to formally model human behavior, which finds application in a wide range of scenarios. In \cite{Hales05:ASI}, the authors propose an algorithm that maintains high levels of cooperation in a peer-to-peer network consisting of entities that perform the collective task of file sharing. The algorithm is adapted from \textit{tag} models of cooperation in social science, which do not rely on explicit reciprocity, reputation or trust mechanisms. Moreover, it does not require any central controller. In \cite{Driggs18:RIH}, the authors present an optimization-based method for approximating the stochastic reachable sets, for informative prediction in human-in-the-loop systems. Reference \cite{Munir14:REW} proposes a human-in-the-loop control mechanism based on human data analysis, which results in a reduction of energy-waste by detection of distractions. In \cite{Ye16:PBS}, the authors consider the repeated task management problem. To include the human in the system, they utilize a model based on \textit{prospect theory}. Reference \cite{Shao19:MCB} investigates the crowdsourcing of multimedia files, where the \textit{cognitive hierarchy} concept is applied to formalize the decision-making of human agents in saving and sharing files. Finally, in \cite{Hu20:ODM}, the authors consider the cyber security problem by formulating the attacker-defender interactions as an evolutionary game. They model the behavior of real-world players using \textit{quantal best-response} dynamics. \\
In \textbf{Table \ref{Tb:SoA}}, we summarize some of the research papers that involve human agent and use specific computational models for human decision-making.
\begin{table*}[ht!]
\small
\centering
\caption{Some Research Works with Human-in-the-Loop\vspace{10pt}}
\begin{tabular}{|p{2cm}|p{5.5cm}|p{9cm}|}
\hline 
\textbf{Reference} &\textbf{Challenge} & \textbf{Approach} \\ \hline \hline
\cite{Hales05:ASI} & File sharing & Modeling human based on tag models of cooperation \\ \hline
\cite{Driggs18:RIH} & Informative prediction of human action & Optimization \\ \hline
\cite{Munir14:REW} & Energy efficiency & Modeling human based on data analysis \\ \hline
\cite{Ye16:PBS} & Repeated task allocation & Modeling human utilizing prospect theory \\ \hline
\cite{Shao19:MCB} & Multimedia crowdsourcing & Modeling human by applying cognitive hierarchy \\ \hline
\cite{Hu20:ODM}& Cyber security defense & Modeling real-world players using quantal response dynamics\\ \hline
\end{tabular}
\label{Tb:SoA}
\end{table*}
\subsection{Application}
\label{sec:App}
The decision-making of humans affects network planning in a variety of applications related to the IoE, especially those involving cooperation or competition. In the following, we provide a few examples.\\
\textbf{Fog Computing:}\\
IoE necessitates energy-efficient and low-latency computation, which is impossible to achieve if the central cloud is the sole option for computation offloading. To address this challenge, fog computing employs smart devices located at the proximity of users for intermediate computation and storage of small- or medium-scale tasks, while the computationally-expensive tasks are still forwarded to the cloud. Given the rapidly increasing number of smart hand-held devices, motivating users to participate in fog computation using their idle devices would result in a significant reduction in deployment, transmission, and computation cost. Moreover, it is possible to pursue users to select fog nodes over the central cloud, or \textit{vice versa}. Therefore, to realize the full potential of fog computing, it is essential to develop appropriate models for human decision-making. \\ 
\textbf{Device-to-Device (D2D) Caching:}\\
The mobile data traffic caused by on-demand multimedia transmission exhibits the \textit{asynchronous content reuse property}; that is, the users request a few popular files at different times, but at relatively small time intervals. Wireless caching takes advantage of this property: Instead of being fetched from the core network frequently in small time intervals, the popular files are stored and re-transmitted whenever requested by new users. Among different models of strategies for wireless caching, D2D caching involves user devices to build a virtual library by storing different files privately and provide each other with the files on-demand. Therefore, similar to fog-computing, it is essential to provide human users with incentives to collect and distribute multi-media contents. \\
\textbf{Opportunistic Cooperative Communications:}\\ 
The basic idea of cooperative communications is that multiple relays cooperate to help a source forward its message to the destination, thereby enhancing the throughput and extending the coverage. In such a scheme, the relays are not pre-determined and pre-deployed. Instead, nearby devices simply act as relays to forward the data. Naturally, acting as relays necessitates an expenditure of the radio resources such as spectrum and power. Therefore, unless a user receives an attractive pay-off, it would not allow its device to cooperate in transmission. Developing suitable reimbursement schemes, however, requires modeling the decision-making behavior of the human. \\ 
\textbf{Wireless Energy Transfer:}\\ 
Wireless energy transfer between devices is a promising idea for the charging of wireless sensor networks, small mobile equipment, on-body medical devices. As the energy waste during the wireless transfer grows with increased distance, it pairs well with energy harvesting. While each device obtains its required energy partially thorough energy harvesting, the rest can be acquired from the nearby devices that do not urgently need energy. Because there is a high density of human-driven devices in many geographical areas, they are considered as suitable choices to participate ins such energy transfers. However, reimbursement is required to justify such a transfer. To set prices, the network manager and automated participants need to model the decision-makers optimally to allocate their budget efficiently and maximize their revenue or minimize their costs.

The performance of every technological network (such as a wireless network) depends on several, often conflicting, variables. Therefore, given the system's characteristics, the network manager designs a specific optimization objective function for each particular application. The optimization sometimes involves model development as well (e.g., using game theory). The model development strongly relies on the problem under investigation while it is essential to also consider the general performance metrics such as complexity and stability. In this regard, human-in-the-loop IoT applications are similar to their counterparts that do not involve human intervention. The difference often arises in formalizing the model taking humans' irrational behavior into account. Some scenarios such as fog computing, wireless energy transfer, and cooperative routing, resemble a market to provide commodities and services. Consequently, the agents require incentives such as monetary compensation. Therefore, to improve efficiency, it is essential to acquire a good understanding of the utility and the behavior of the involved human decision-makers, for example, by suitably adapting the utility functions; otherwise, appropriate mechanism design is implausible. Some other scenarios such as D2D caching or autonomous driving require coordination. In such cases, it is crucial to predict the actions of human decision-makers, utilizing appropriate models or possibly using its features, similarity measures, and historical actions. In summary, the unrealistic assumption that all decision-makers are fully rational is a pitfall of traditional models that shall be addressed.
\section{Agent-Based Models of Human Behavior}
\label{sec:HumanModel}
Here we describe the most important models of human behavior. These models aim to predict human behavior in scenarios that can be modeled by a game-theoretic framework. Contrary to typical machine learning methods, they can operate with no historical data about human behavior in games identical to the current game, so long as they have access data about similar agents' choices in similar games. This data is used to estimate parameters in behavioral models. The models aim at adapting the classical game theory to capture real human behavior. To achieve this goal, they combine the standard results with insights from cognitive- and behavioral psychology and microeconomics. These models seek to predict behavior in one-shot, simultaneous games where participants do not learn (whether due to one-off participation or since the dynamics of the game change sufficiently fast to render learning useless). Indeed, almost all of the theoretical justification from behavioral psychology for these models is based on the one-off interaction, as further interactions may be complicated by the relationship between agents. This means that if available strategies in a system are not changing over time, then as data is accumulated from observation of agents, a learning approach can be used to improve or adjust the original general model of human behavior. The usefulness of the models below, in such a static system, would be then either to 'start-off' such a prediction model, or to use as a baseline for performance evaluation.
\subsection*{A Brief Background in Game Theory}
\emph{Normal-form} games can be expressed as a matrix of payoffs for players, dependent on the combination of choices made by each player, or, in other words, the \textit{joint action profile}. Each possible action for a player is referred to as a \textit{pure strategy}. A \textit{mixed strategy} is then a probability distribution over pure strategies. The \textit{expected utility} of a pure strategy for a player is the expectation of the payoff of a particular action, given the \textit{belief} of that player about the opponents' joint action profile. 

In classical game theory, any \textit{rational} agent seeks to maximize its expected utility. Human agents, however, often act irrationally due to emotions, social norms, peer pressure, and the like. The experiments leading to the development of human models for such games usually involve showing such a normal form depiction to experiment participants and asking them to make a choice. In this paper, we describe computational models for bounded-rational or irrational actors, who we assume in general are selfish, but due to either emotions, biases or limitations on human cognitive ability, are not optimal decision-makers. Note that although mixed strategies play a crucial role in classical game theory, they will not appear so often in these models. The probability distributions over actions that appear here indicate the varied levels of human cognition that our models seek to capture rather than being a result of players' strategic thinking. 
\subsection{Iterative Strategic Thinking}
\label{subsec:Iterative}
The concept of Nash equilibrium is based on the assumption that agents reason their way to an equilibrium preemptively before taking any actions, via \textit{fixed-point reasoning} in which the agents find some equilibrium from which a unilateral deviation does not improve the utility. A natural alternative is to learn and develop towards such an equilibrium, where the agents have the opportunity to refine their strategy over time. In the absence of such an opportunity for learning, we are left with \textit{iterative strategic thinking}, in which an agent reasons along the lines of 'she thinks that I think that she thinks' before making any conclusion. A natural example is a financial marketplace, in which participants price a stock by guessing at how other participants will value it in the future. 
\subsubsection{Level-$k$ Reasoning}
\label{subsubsec:Level}
Human agents participate in a limited number of iterations of strategic reasoning. A level-$k$ agent engages in $k$ iterations, responding to level-($k$-1) agents by recursively determining their behavior, and responding to it by selecting the action maximizing their expected utility \cite{Wright2017}. The simplest model for the level-$0$ agents is uniform randomization over available actions, and more complex specification is discussed below. The final model is a linear combination of $K$ agent levels parameterized by $K$ variables.
\subsubsection{Cognitive Hierarchy}
\label{subsubsec:Cognitive}
A subtle feature of the level-$k$ model is that each agent considers all of the other agents to be one level below itself in the cognitive hierarchy. Alternatively, they can respond to a linear combination of the strategies of lower agents in the cognitive hierarchy model \cite{Wright2017}. This can be specified in several ways. Each level can have its own belief about the proportion of lower-level agents (introducing several parameters), or these distributions can be the actual distribution of agent levels of the model, normalized appropriately. For an agent at level-$K$, a popular way to specify the distribution of agents is to assume that the number of agents in each level $k=1,...,K-1$ follows a Poisson distribution. 
\subsubsection{Specifying Level-0 Behavior}
The distribution of level types in the level-$k$ or cognitive hierarchy model significantly affects the proper specification of level-0 behavior. 

If there is no way to capture the errors made by level-$k$ players (see also Section \ref{subsubsec:Quantal}), then the level-0 specification captures such errors by a simple behavior following uniform randomization as previously described. Level-0 behavior can also capture the salient features of a particular setting or commonly observed heuristics used by strategically naive players. See \cite{Crawford2013, Wright2014} for a summary of how modifications to level-$k$ and cognitive hierarchy models improved models for games with asymmetric information, market entry and other coordination games, games in which salient features affect coordination efforts, and games of strategic communication.
\subsection{Alternatives to Best-Response}
\label{subsec:AltBest}
In a perfect-rationality setting, best-response dynamics, which return Nash equilibria if they exist, are unrealistically deterministic in their output. Part of the heterogeneity of human responses to situations can be explained by their different cognitive abilities and methods of problem-solving, as with the level-$k$ and cognitive hierarchy models above. Another part can be explained as people simply making mistakes - even if they are attempting to act strategically. The models below incorporate \textit{imprecision} into a \textit{noisy best-response}. 
\subsubsection{$\epsilon$-Nash}
\label{subsubsec:Epsilon}
The simplest such noisy best-response function is to return a pure Nash equilibrium (if any exists) with some probability $1-\epsilon$, and with probability $\epsilon$, uniformly randomize over the remaining actions \cite{Wright2017}. However, this approach has been shown to poorly model individuals' behaviour, mirroring the observed fact that people make \emph{systematically} rather than \emph{randomly} irrational decisions.

We use $\epsilon$-Nash in the example in Section \ref{sec:Example} to simply demonstrate the effect of a small amount of noise on the behaviour of agents. If we wanted to model agents' behaviour accurately, then we would use more sophisticated models, such as QBEs, described in the next section.
\subsubsection{Quantal Best-Response}
\label{subsubsec:Quantal}
Quantal Best-Response (QBR) returns a probability distribution over actions, satisfying two observations from behavioral psychology: (i) An agent is more likely to choose an action with greater expected value than an action with less expected value; (ii) The likelihood of making the optimal choice decreases as the importance of the decision decreases, whether because the difference between actions decreases, or because the payoffs decrease in value \cite{McKelvey1995}. The most common variation is the logit-QBR function, which selects an action $s_{i}$ with higher probability if, under the belief $B_{i}$, it has a higher expected utility $U_{i}(s_{i},B_{i})$:
\begin{equation*}
\text{QBR}_{i}(s_{i},B_{i},\lambda)=\frac{\exp({\lambda U_{i}(s_{i},B_{i}))}}{\sum_{s'_{i} \in S_{i}}\exp{(\lambda U_{i}(s'_{i},B_{i}))}}
\end{equation*}
The precision parameter $\lambda$ corresponds to the rationality of agents. As $\lambda \to \infty$, QBR approximates best-response, and if $\lambda \to 0$, QBR approaches uniform randomization.

This can be used on its own, predicting the \emph{quantal best-response equilibrium} via similar fixed-point reasoning to that involved in the calculation of Nash equilibria. However, this reasoning is similarly (or more) difficult than even Nash equilibria calculation. QBR can nonetheless be incorporated into level-$k$ and other models, replacement of the previous utility maximization response. Incorporating it usually improves predictive performance at the cost of introducing only one more parameter-although variants are possible, such as different precision levels for different agent types \cite{Wright2017}. 
\subsubsection{Noisy Introspection}
\label{subsubsec:Noisy}
An alternative way to incorporate QBR is noisy introspection, in which agents have no limit on the number of strategic iterations they can perform, but their reasoning becomes noisier as they perform higher levels of introspection, with a corresponding exponential increase in the noise parameter \cite{WRIGHT201716}.
\subsection{Adjustments to utility function}
\label{subsec:Asjust}
The first group of models, iterative strategic thinking, offer alternatives to the fixed-point reasoning used in classical game theory. The second group of models introduce systematic models of noisiness in decision-making. The third group, described here, corresponds to changing the underlying utility function, which produces the final payoffs for agents, which their decisions are based upon.
\subsubsection{Prospect Theory}
\label{subsec:Propspect}
A highly popular explanation for many non-rational behaviors commonly exhibited by human decision-makers is \emph{prospect theory}. It is an alternative to conventional utility maximization in decision theory, emphasizing people's biased perceptions of value and probability. As such, there are two main areas of adjustments: how the agent perceives the values of outcomes (value function), and how the agent perceives the probabilities of outcomes (weighting function).
\paragraph{Value Function}
\label{subsubsec:Value}
In prospect theory, there are three key changes to way outcomes are valued. Firstly, a possible outcome of an action, rather than having a fixed 'utility' value, is seen as a `gain' or `loss' relative to their \emph{reference} state, or current state. Secondly, the base gain or loss of an outcome is adjusted by a \emph{value function}, following the notion of `diminishing returns' and `diminishing losses' with the increase in value of the shift. Thirdly, the value function for gains is scaled-down relative to that of losses, so an agent will perceive the absolute value of a loss as larger than that of a gain, reflecting people's psychological aversion to loss.
%
\paragraph{Weighting Function}
\label{subsubsec:Weighting}
Events' probabilities are also transformed via a weighting function. In the presence of uncertainty, each outcome is associated with a relative probability of occurrence $p_i$. The weight function maps from the probability $p_{i}$ of an outcome $x_{i}$ to a point in $[0,1]$, though its output is not a probability measure. The weighting function reflects people's tendency to exaggerate the difference between complete certainty and slight uncertainty. Among several others, \cite{tversky1992} provides possible parametric functional forms for this relationship. The value of parameters can be learned from the data. Caution is recommended where learned parameters' values lead to nonsensical weighting functions, such as inverting the curvature. 

Finally, we describe some limitations of the prospect theory. Since prospect theory focuses on adjusting agents' utility functions, we might consider combining it with models of iterative decision-making, such as level-$k$. However, this would create tension between the assumptions of iterative strategic thinking and that of prospect theory. Namely, an agent in iterative strategic thinking models others using the same utility function which they use, whereas Prospect Theory is based on psychological research that suggests that agents are unaware of the biases present in their own and others' reasoning. There is no obvious satisfactory resolution of this tension, and exploring the applicability of prospect theory to game-theoretic situations is an interesting direction for research.
\subsubsection{Social Preference}
\label{subsec:Social}
A large class of models seek to explain non-rational behaviour via \textit{social preferences} \cite{social_prefs1}. Such models formalize notions such as aversion to inequity and reciprocity. The formalization is then utilized to predict human behavior in situations where conventional game theory incorrectly predicts strongly selfish behavior. In these models, rather than modelling agents as noisily deviating from best-response, the utility functions of agents are given by a linear combination of their normal or base `selfish' utility functions, and terms that incorporate notions of social utility. By altering the weighting of the different components of the utility function, either selfish behaviours, or altruistic or other `non-rational' behaviours, can be emphasised.

There are then many possible behaviours which can be modelled via using a utility function. One example is \emph{altruism}, in which the utility function increases with global base utility. Another example is \emph{inequity aversion}, in which the utility function has a term which decreases with inequity. Inequity can be measured in a number of ways, one of which is simply to measure the difference between the largest and smallest base utilities. Another possibility is to compare the utility of each individual with the average utility of the society. We can also model `negative' behaviours. An example of this is envy, which is modelled via a term in their utility function. This term decreases with the difference between their base utility and that of the highest-utility agent.

Notably, this does not require any complex iterative procedures. A clear difference between this method and the previous methods is that the `noisiness' is not a result of players making mistakes, but rather playing best-responses according to notions such as what is fair or unfair for themselves or society. As such, they model significantly different aspects of agents' behaviour. This difference should be taken into account when modelling. 

In \textbf{Table \ref{Tb:ModelIoT}}, we summarize the well-known models of human behavior together with some of their application areas in IoE.
\begin{table*}[ht!]
\small
\centering
\caption{Well-known Models of Human Decision-Making with IoT Application Areas \vspace{10pt}}
\begin{tabular}{|m{4.3cm}|m{5cm}|m{7.2cm}|}
\hline 
\textbf{Model} &\textbf{Potential Application Area} & \textbf{Brief Description} \\ \hline \hline
Iterative strategic thinking & 
Autonomous driving, E-commerce, D2D caching & Making decisions based on some beliefs on how others will decide, rather than the hypothetical reasoning of game theory .\\ \hline
Alternation to best-response &
 Cooperative communications, Fog computing,  Smart grid& Making decisions to optimize an objective function such as when bidding, but suffering from mistakes. \\ \hline
Adjustment to utility function & 
 Sharing economy, Crowdsourcing, Wireless energy transfer, Cooperative mining & Incorporating factors beyond rational self-interest, such as loss aversion (prospect theory) or a desire for fairness (social preferences). \\  \hline
\end{tabular}
\label{Tb:ModelIoT}
\end{table*}
\section{Developing a Model}
\label{sec:DevModel}
As observed in Section \ref{sec:HumanModel}, most of the models that formalize the human's strategic behavior include some parameters. Often, such parameters are optimally tuned via training the model using the available data. Indeed, one of the key features of a behavioral model is its data-driven methodology: rather than normatively defining mathematical features of games, it seeks to predict the decision-making behavior based on past evidence of such behavior. In the following, we provide a brief explanation without going into details. 

One conventional approach for training a model, i.e., to find the parameters' value, is to use the \textit{maximum likelihood estimation} (MLE). The `likelihood' describes the plausibility of a set of parameter values of a statistical model describing a set of observations, given those observations. In an MLE model, there is a  likelihood function associated with a statistical model. The function takes in the vector of parameter values of that model and a data set. The output is then the likelihood of the specified parameter values given the data. More precisely, initially, a model is created using the given vector of parameter values. This model returns a probability distribution over actions. Then, for every element in the training data, it returns the probability of observing the selected action and multiplies these probabilities together. The resulted value is the likelihood that those parameter values describe the observed data. The final parameter values for a given training set are those which maximize the likelihood function. The data required for such model training is usually collected by performing experiments; however, often the usefulness of data from experiments as generalizable training data is obscured by the experiment's design to answer specific questions about applications of behavioral game theory to a particular area. Moreover, the quality of data plays a critical role. For example, self-reported data is frequently biased and inaccurate.

Finally, it is essential to evaluate the developed human decision-making models. Here, the predictive ability of the model is of specific importance. Therefore, such a model should be tested on data separated from that it is trained on, a concept which is known as \textit{cross validation}. Other important factors are low-complexity, interpretability, and accuracy. Often, there is a trade-off between these factors. For example, using the likelihood as a measure of accuracy, despite being simple, lacks interpretability to some extent, as it does not give much information about the importance of different model parameters.  
\section{Example Application: Computation Offloading in Fog Computing}
\label{sec:Example}
Here we provide an example concerning fog computation offloading to show the effect of humans' irrational or imprecise behavior on the system's performance.  
\subsection{System Model and Problem Formulation}
\label{subsec:SysModel}
We consider a scenario with an offloading user that divides  some dividable tasks between a set of computing fog nodes represented by $m \in \mathcal{M}=\{1,2,...,M\}$. Each fog node demands a price $c_{m}$ to perform each unit of the task. We denote the share of each fog node $m \in \mathcal{M}$ by $r_{m} \in \mathbb{R}^{+}$. The user has limited budget so that it should satisfy $\sum_{m\in \mathcal{M}}r_{m}c_{m}\leq B$, with $B>0$ being the budget. 

We cast the task allocation problem as a virtual `supply-demand market' by designing a two-level game to describe the interactions among fog nodes and the offloading user. On one side, the fog nodes represent the sellers that set the service prices. On the other side, the offloading user represents the consumer that uses the services by offloading the task at the given prices. Therefore, the user(s) and the fog nodes follow the iterative process summarized in \textbf{Alg. \ref{alg:SuplyDemand}}.  
\begin{algorithm}
\caption{Task Offloading to Fog Nodes}
\label{alg:SuplyDemand}
\small
\begin{algorithmic}[1]   
\STATE Each fog node $m \in \mathcal{M}$ announces the initial price $c_{m}$;
\REPEAT{
\STATE The user determines the demand $r_{m}$, $m \in \mathcal{M}$;
\STATE Each fog node $m \in \mathcal{M}$ adapts the price $c_{m}$.
}
\UNTIL{convergence}
\end{algorithmic}
\end{algorithm}

The performance of the offloading user depends on several crucial factors, including (i) transmission delay and service delay; (ii) power consumption for data transmission; and (iii) paid service price to the fog nodes. We define the utility function of the offloading user as
\begin{equation}
\label{eq:UtilUser}
g_{o}(r_{1},...,r_{M})=a\sum_{m \in \mathcal{M}} \alpha_{m} \log(r_{m}\beta_{m})-c_{m}r_{m}.
\end{equation}
In (\ref{eq:UtilUser}), the first term is the utility of the user by offloading the task. We model the utility by an increasing concave function (e.g., logarithmic function) due to the following reason: Although offloading is in general beneficial for the user, the marginal utility of offloading decreases since heavier offloading results in longer delays and higher energy consumption. The parameters $\alpha_{m}$ and $\beta_{m}$ depend on the characteristics of the fog node that affect the user's quality of experience, for example, the distance to the offloading user, queue length, and the like. Detailed description of such dependencies is out of the scope of this paper. The second term in the right-hand-side of (\ref{eq:UtilUser}) is the service cost paid to the fog nodes $m \in \mathcal{M}$. The gain is then the utility minus cost.

Similar to the offloading user, each fog node $m \in \mathcal{M}$ has some measure for its gain. The utility of fog node $m$ equals the price paid by the offloading user. The cost is then determined by several factors such as the consumed energy. We model the cost of a fog node $m$ by a linear function with parameter $c_{m}^{(\text{L})}$. Formally, the gain of a fog node $m \in \mathcal{M}$ yields
\begin{equation}
\label{eq:UtilFog}
g_{m}(c_{m})=\kappa_{m}(c_{m}r_{m}-c_{m}^{(\text{L})}r_{m}),
\end{equation}
where $\kappa_{m} \in (0,1]$ is a price-regulatory constant, which is equal to one in the most conventional form. Naturally, $c_{m} \geq c_{m}^{(\text{L})}$. The traditional solution of the supply-demand procedure described in \textbf{Alg. \ref{alg:SuplyDemand}} is the `Nash equilibrium', as determined in the following section.
\subsection{Nash Solution}
\label{subsec:NashSolution}
At Nash equilibrium, the players act by best-responding to each other. Indeed, Nash equilibrium is the intersection of the best-response dynamics of the players. Therefore, no player benefits by a unilateral deviation from the equilibrium strategy, implying a steady state.  

In the developed offloading game, the offloading user, if fully rational, optimizes its performance by maximizing the utility function (\ref{eq:UtilUser}) subject to the constraint $\sum_{m\in \mathcal{M}}c_{m}r_{m}=B$. This problem is an equality-constrained convex optimization problem with an one-dimensional differentiable function for which `Karush-Kuhn-Tucker' (KKT) conditions hold. It requires only a few calculus steps to show that $r^{*}_{m}=\frac{B}{c_{m}(1+\sum_{i \in \mathcal{M},i \neq m}\frac{\alpha_{j}}{\alpha_{m}})}$.

Similarly, every fully-rational fog node $m \in \mathcal{M}$ determines the service price by maximizing the utility function given by (\ref{eq:UtilFog}) that corresponds to the best-response. As the equilibrium is the intersection of the two best-response dynamics, the optimal price is the solution of the following $r^{*}_{m}+\kappa_{m}(c_{m}-c_{m}^{(\text{L})})\frac{\partial r^{*}_{m}}{\partial c_{m}}=0$.

In the distributed implementation, the fog nodes start from some minimum price and adapt the prices continuously until convergence, i.e., until the price and demand settle at some point. Based on an approach similar to the one that appears in \cite{Maghsudi11:AHC}, one can establish that with the given price and demand functions, the procedure described in \textbf{Alg. \ref{alg:SuplyDemand}} indeed converges to an stable point that is the Nash equilibrium of the formulated game. 

For simulation, we consider a system with one offloading user and four fog nodes. \textbf{Fig. \ref{Fig:Rational}} depicts the iterative interaction between the fog nodes and the user. It is obvious that the price, and thus the demand, settle at some point. The characteristics of this settlement point mainly depend on the preferences of the user. These, in turn, are determined by the properties of the fog node, including their requested service price.
\subsection{The Effect of Human Agent with Noisy Best-Response}
\label{subsec:Human-Agent}
In fog computing, similar to other crowd sourcing schemes, the set of contributors might include small hand-held devices, where the human agents take the role of decision-makers. To model the human agents in the aforementioned fog computing scenario, we use a noisy best-response model. We model the imprecision of decision-making by a uniform noise that is added to the best-response as perturbation. We perform the experiment by two small noise levels, namely maximum of 5\% and 10\% of the best-response for all fog nodes and the offloading user. \textbf{Fig. \ref{Fig:NoisyLow}} and \textbf{Fig. \ref{Fig:NoisyLarge}} depict the results respectively. From the figures, it is obvious that even a small deviation from the fully-rational response can result in dramatic instability in the system, as well as prolonging the convergence time; As such, it should be taken into account for network optimization.    
\subsection{Potential Solution to Resolve the Instability}
\label{subsec:Stability}
\begin{figure*}[h]
	\centering
	\begin{subfigure}{0.48\textwidth}
	\centering
		\includegraphics[width=\linewidth]{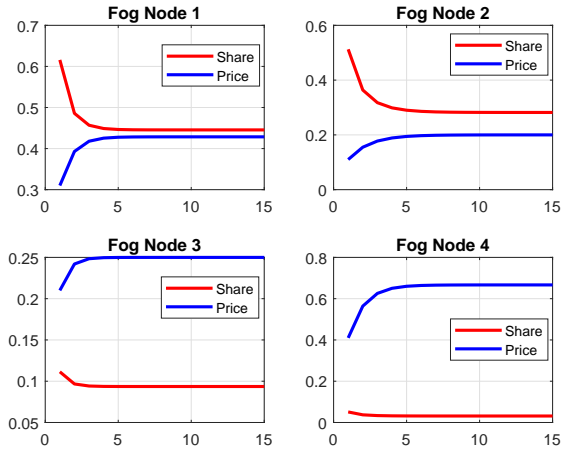}
		\caption{Fully rational agents.}
    \label{Fig:Rational}
	\end{subfigure} \ \
\medskip	
	\begin{subfigure}{0.48\textwidth}
	\centering
		\includegraphics[width=\linewidth]{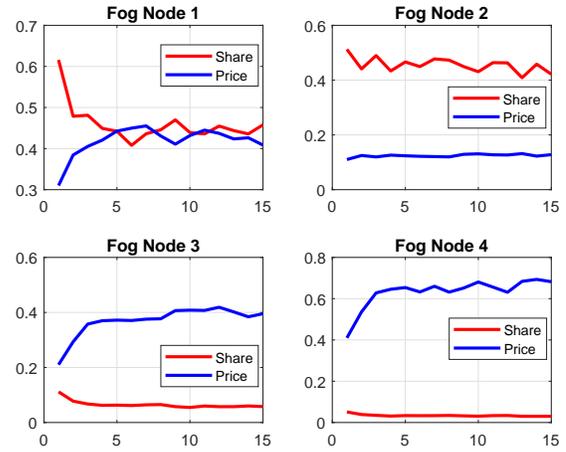}
		\caption{Bounded-rational agents with 5\% noisy best-response.}
    \label{Fig:NoisyLow}
	\end{subfigure} \ \
\medskip
	\begin{subfigure}{0.48\textwidth}
	\centering
		\includegraphics[width=\linewidth]{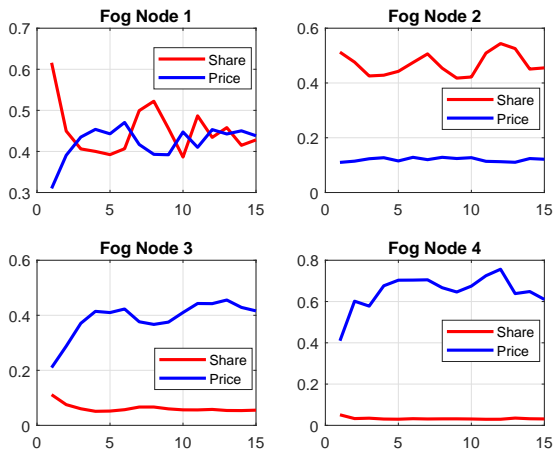}
		\caption{Bounded-rational agents with 10\% noisy best-response. }\ \
    \label{Fig:NoisyLarge}
	\end{subfigure}
\medskip
   \begin{subfigure}{0.48\textwidth}
   \centering
	\includegraphics[width=\linewidth]{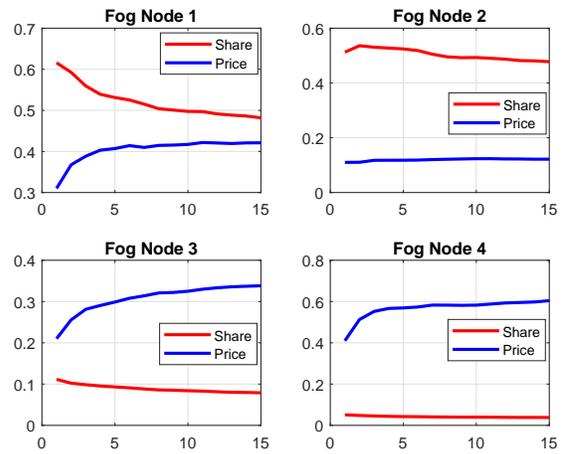}
	\caption{Approximate resolution of instability.}\ \
   \label{Fig:Stability}
  \end{subfigure}
\medskip
\caption{Price-demand negotiation. The x-axis shows the iteration rounds. Prices are normalized by 10. The resolution of instability results from signal averaging over 10\% noisy best-response.}
\label{plots}
\end{figure*}

To resolve the instability of the system that is induced by the noisy best-response, we use a simple, yet effective, signal processing approach, namely, \textit{signal averaging}. Utilizing this method relies on the assumption that every player is aware of the possibility of making mistakes; therefore, at every round, each player simply acts by the moving average of its best-responses in all the pricing-demand rounds so far. As it is evident from \textbf{Fig. \ref{Fig:Stability}}, this simple method reduces the noise significantly and resolves the instability to a large extent. 

\begin{figure}[t]
\centering
\includegraphics[width=0.49\textwidth]{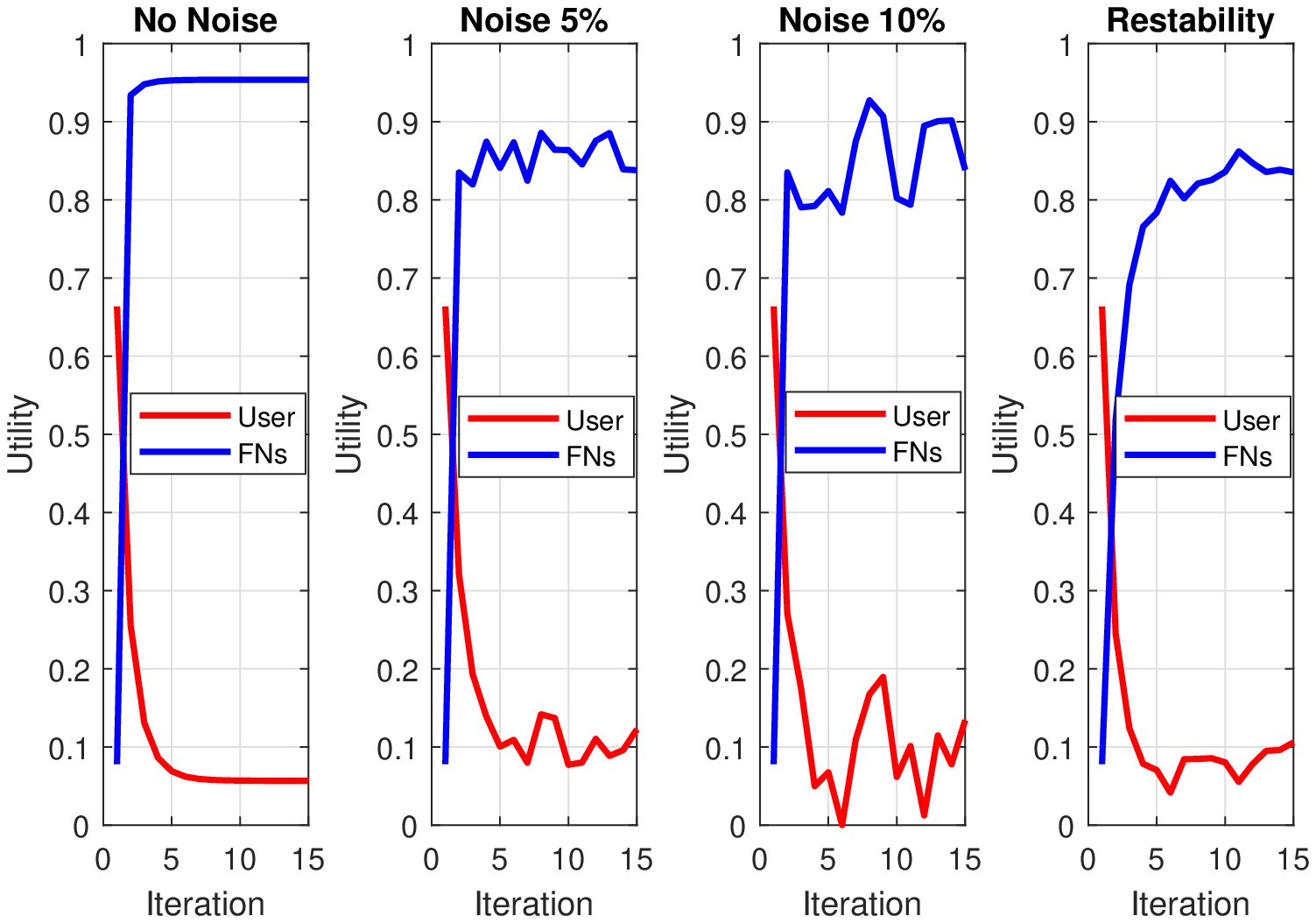}
\caption{Utility of the offloading user and the aggregated utility of the fog nodes at full rationality, in different noise levels, and when using the signal averaging method.}
\label{Fig:Utility}
\end{figure}

Finally, \textbf{Fig. \ref{Fig:Utility}} shows the utility of the offloading user (as the customer) and the aggregated utility of the fog nodes (as the suppliers) in different scenarios. It can be concluded that in addition to system stability, impreciseness reduces the system efficiency.
\section{Outlook and Conclusion}
\label{sec:Conc}
Humans play a vital role in future IoT networks. As such, appropriate modeling of human behavior is significantly important for network optimization. We elaborated on formalizing human behavior using simple and efficient, yet effective, models. Based on a toy example in the fog computing framework, we showed that neglecting the human-specific characteristics endangers the system's stability. Therefore, in future-looking applications, the models of wireless communication networks based on multi-agent systems with full-rationality shall be reconsidered to include the imperfections in human behavior. 

The complexity of humans' character and behavior, however, render such enhancement notoriously difficult. While such difficulties have been studied in several domains such as marketing, the available solutions are not directly generalizable to technological networks such as IoT, and further study is imperative. Some challenges are as follows:
\begin{itemize}
\item Generally, one uses the available data to tune the parameters of the computational models; however, IoT applications are widely diverse. Moreover, gathering a large amount of data might not be plausible. As such, it is necessary to develop some methods to manually adapt and re-use the existing data for different scenarios, or to create synthetic data. 
\item It is important to adapt the standard human decision-making models to the IoT applications, as such applications are influenced by various physical constraints, e.g., the transmission medium over which the entities communicate with each other.
\item Centralized learning can be excessively time- and energy-consuming; hence it is essential to develop distributed learning algorithms to obtain accurate models and parameters efficiently and securely.
\item Another challenge is the implementation of human-in-the-loop solutions in existing IoT systems. Indeed, the efficiency and feasibility of such implementation is itself a line of research, as it might require large computing capability and availability of power resources.
\item Investigating the effect of integrating humans in the networks on addressing the trade-offs in IoT applications, as well as on network optimization and management. 
\end{itemize}
Beyond IoE applications and in the general setting, there are several challenges open to address. These include 
\begin{itemize}
\item As previously discussed, there are several models to computationally formalize human decision-makers. Therefore, it is necessary to develop accurate and application-specific indicators for the complexity and permissiveness or usefulness of probabilistic models.
\item Experiment design is a classical research direction in behavioral studies. In this regard, it is important to design diverse tests that reflect the usability of the combination of two or more models. 
\item Although an entity might change its behavior over time, such dynamics are largely neglected in the current research. Therefore, it is imperative to study models that include the time-variations in the behavior of a human agent.
\item Irrespective of the nature of players, efficiency and stability are the most important factors in autonomous systems. Therefore, developing methods to eliminate the adverse effect of human agents on the system's stability is a significant line of research.
\item Tailoring deep learning frameworks for an accurate estimation of the models' parameters is another potential research direction. 
\end{itemize}
\bibliographystyle{IEEEbib}
\bibliography{Main}
\end{document}